# An AI Based Super Nodes Selection Algorithm in BlockChain Networks

Jianwen Chen[1], Kai Duan[2], Rumin Zhang[1*], Liaoyuan Zeng[1], Wenyi Wang[1]

1. MediaLab, University of Electronic Science and Technology of China, Chengdu 611731

2. AICHAIN Foundation Ltd. Singapore

## Abstract

In blockchain systems, especially cryptographic currencies such as Bitcoin, the double-spending and Byzantine-general-like problem are solved by reaching consensus protocols among all nodes. The state-of-the-art protocols include Proof-of-Work, Proof-of-Stake and Delegated-Proof-of-Stake. Proof-of-Work urges nodes to prove their computing power measured in hash rate in a crypto-puzzle solving competition. The other two take into account the amount of stake of each nodes and even design a vote in Delegated-Proof-of-Stake. However, these frameworks have several drawbacks, such as consuming a large number of electricity, leading the whole blockchain to a centralized system and so on. In this paper, we propose the conceptual framework, fundamental theory and research methodology, based on artificial intelligence technology that exploits nearly complementary information of each nodes. And we designed a particular convolutional neural network and a dynamic threshold, which obtained the super nodes and the random nodes, to reach the consensus. Experimental results demonstrate that our framework combines the advantages of Proof-of-Work, Proof-of-Stake and Delegated-Proof-of-Stake by avoiding complicated hash operation and monopoly. Furthermore, it compares favorably to the three state-of-the-art consensus frameworks, in terms of security and the speed of transaction confirmation.

## 1. Introduction

Bitcoin[1], one of the most well-known cryptocurrency[2], has demonstrated that it is practical and valuable to use the blockchain as a transaction ledger. Essentially, the blockchain is a distributed database of public ledger of all transactions or digital events that have been executed and shared among participating parties [3]. Its most notable characteristics are its decentralization and collective maintenance. Besides, time-series data, programmability and security are meaningful properties of blockchain as well. Furthermore, as a universal underlying technology framework, blockchain systems can be successfully applied to other financial and even non-financial applications.

Every blockchain system, especially cryptographic currencies such as Bitcoin, always face a latency-confidence trade-off in a transaction [4]. Since a higher confident transaction requires a longer time to be confirmed by nodes, those applications which require low latency cannot ensure whether their transactions will be confirmed. Meanwhile, the system needs to prevent double-spending and solve a problem similar to the Byzantine general problem [5]. Blockchain works out the two problems by reaching consensus protocols among all nodes. How to reach consensus more efficiently and correctly is a core issue and technology in the blockchain. There are several representative methods recently, including Proof-of-Work (*PoW*), Proof-of-Stake (*PoS*) and Delegated Proof of Stake (*DPoS*).

Bitcoin addresses the consensus problem using PoW, which urges nodes to repeatedly compute hashes to grow the blockchain. It is a way of getting compensation by working and can achieve complete decentralization, free access to nodes, and avoid the cost of establishing and maintaining a centralized credit institution. Nevertheless, adopting this scheme will result in a large amount of waste of resources, and will also lead to a hig h concentration of computing power, deviating from the original intention of the decentralized design. At the same time, the mechanism takes a long time to reach a consensus and is therefore not suitable for commercial applications.

*PoS* tries to find authoritative nodes by the stakes of the original cryptocurrency, which partly solve the drawbacks of *PoW*. But the system is easily monopolized by nodes with large stacks and unfair for new participants. Try to settle these problems, a new method called Delegated Proof of Stake (*DPoS*) allows each node to vote, thereby generate a certain number of representatives, whose rights are exactly equal to each other. Whereas this method cannot be suitable for a completely decentralized scenario. In addition, in the scenario where the number of network nodes is small, the representative of the authoritative nodes is not strong.

The waste of resources, inverse decentralization process and long time to reach consensus are three key problems of the consensus protocol. The above three methods, as well as others, which including Practical Byzantine Fault Tolerance (*PBFT*), cannot solve or only partially solve these problems.

In this paper, in order to solve these problems, we propose a novel nodes selection algorithm based on AI technology. First, the average transaction of each node is calculated by AI algorithm. Then, we make statistics of threshold values of average transaction number of nodes. According to the threshold values, all mining nodes in the blockchain networks are divided into three categories: super nodes, random nodes and validator node. The main contributions of our work are as follows:

(1) To save resource, avoid the complicated hash operation and redundant verification operation are avoided.
(2) Considering the security, the random mining nodes selection strategy is employed for realizing the decantation.
(3) A node capability mechanism is developed which will shorten the cycle of reaching the consensus.

The remainder of this paper is organized as follows. Sec 2 reviews related works on the core technology about blockchain and the existing method to combine the AI with blockchain. The details of our proposed algorithm are elaborated in Sec 3. Sec 4 shows the experiments and Sec 5 concludes this paper.

## 2. Related Work

### 2.1 Fundamental technology about blockchain

In recent years, a great amount of researches have been conducted in distributed data storage, point-to-point transmission, consensus mechanism, encryption algorithm and other computer technologies.

Guilford J D [6] proposed the SHA256 algorithm which is employed in the blockchain, namely the original transaction record of any length is computed twice by SHA256 algorithm to obtain the hash value and the hash value's length is 256.The following Merkle tree and *POW* are the applications of hashing algorithm. In the blockchain, the Merkle tree [7-9] is utilized to store transaction information and generate the digital signature of the transaction set. Merkle tree greatly improves the efficiency and scalability of the block chain. Then, the Merkle tree can also verify the data without running the complete block chain network node.

In order to solve the problem of "doublespending", timestamp [10,11] was introduced to record the writing time of the block data, making it possible for data to reconstruct the history. In addition, as proof of existence, timestamp ensures that blockchain's database is not tampered with and forged.

P2P technology is used to make each node on the network communication have the equal status, and there is no specialized center node and hierarchy structure. Each node will undertake the network routing and data validation, data transmission. In order to realize the security of data transmission and ownership verification, the blockchain uses the asymmetric encryption algorithm called ECC (Elliptic Curve Cryptograph), and each user has a pair of keys, one public and one private [12]. Users sign the transaction information with ECC, meanwhile other users can verify the signature with the public key of the signed user. In addition, the public key is also used to identify different users and construct their bitcoin addresses.

## 2.2 Proof-of-Work (POW)

*PoW* was initially introduced by C. Dwork and M. Naor to combat email spam. *PoW* was proposed to mitigate distributed denial-of-service attacks [13]. Nakamoto Satoshi adopted the *PoW* method in Bitcoin systems [14].

Bentov, I etal.[15] presented a hybrid protocol that relies both on *PoW* and Proof of Stake, where the objective is to combine to advantageous properties of the *PoW* element and the Proof of Stake element into a system that is superior to relying on only one of these two elements. Ateniese, G etal.[16,17] proposed an alternative to *PoW* that is based on data storage. Arthur Gervais etal.[18] introduced a novel quantitative framework to analyse the security and performance implications of various con-sensus and network parameters of *PoW* blockchains. they devise optimal adversarial strategies for double-spending and selfish mining while taking into account real world constraints such as network propagation, different block sizes, block generation intervals, information propagation mechanism, and the impact of eclipse attacks. Alex Biryukov etal.[19]construct an asymmetric proof-of-work (*PoW*) based on a computationally-hard problem, primary introduced Equihash, is a *PoW* based on the generalized birthday problem and enhanced Wagner's algorithm for it.

## 2.3 Proof-of-Stake (POS)

The *PoS* method was initially used in Peercoin in 2012. Generally speaking, proof-of-stake means a form of proof of ownership of the currency. unlike the *PoW* method, *PoS* method does not have mining that uses computing power. *PoS* is one of the main candi-dates to solve the energy demand problem in the current blockchain protocols such as Bitcoin and Ethereum[20].

Each party has a certain amount of stake in a blockchain , typically the amount of cryptocurrency. There is a randomized leader election process for each block; and the elected party can release the next block. The more stake a party has, the more likely it is to be elected as a leader. Similarly to *PoW*, block issuing is rewarded. The *PoS* method does not use too much computing power, it is more cost effective than the *PoW* method. However, The fact that a person with a large stake can easily monopolize, it unfair for new participants. This is one of major disadvantages in the *PoS* method. Yuefei Gao etal.[21] proposed a new consensus protocol based on sharding and proof of stake. The scalability of the method is expected to increase linearly with the network size. Fahad Saleh [22] provided the first formal economic model of *PoS* and demonstrate that *PoS* induces consensus in equilibrium.

## 2.4 Delegated Proof-of-Stake（DPOS）

*DPOS* is a new consensus algorithm based on *POW* and *POS* to guarantee the security of digital currency network. *DPOS* [23] is a fast, excellent, decentralized, and convenient consensus model. In order to solve consensus problems in a reasonable manner, *DPOS* utilizes the energy to stakeholder endorsement voting. Each holder of the currency can vote, resulting in a certain number of representatives, or a certain number of nodes or mines, and their rights are exactly equal to each other. Holders can change these representatives at any time to maintain the "long-term purity" of the chain system.

The advantage of *DPOS* is that it can minimize the energy consumption of maintaining network operation and manage the operation of the entire chain in a low-cost way, which largely solve the problem of excessive energy consumption caused by *POW* in the mining process. Meanwhile, *DPOS* decentralizes the decision to operate the blockchain network to the nodes of the entire network, avoiding the bias of "trust balance" caused by the distribution of *POS* equity.

## 2.5 Practical Byzantine Fault Tolerance (PBFT)

The Practical Byzantine Fault Tolerance (*PBFT*) [24] is an algorithm for solving a Byzantine Fault resulting from a failure in building a consensus caused by the BGP. It was introduced by Miguel Castro and Barbara Liskov, which solves the problem of low efficiency of the original Byzantine fault tolerant algorithm. *PBFT* is a message-based consistency algorithm that achieves consistency in three phases that may be repeated due to failure. Specifically, it depends on three rounds of message exchange before reaching agreement. This ensures that 3f + 1 nodes can achieve consensus also in presence of f Byzantine nodes [25].

*PBFT* consensus efficiency is high, enabling high-frequency trading. Nevertheless, the entire amount of nodes need be known, and the maximum amount of illegal nodes needs to be set. These requirements make it hard to employ this algorithm with regard to public systems. It was considered to be a serious challenge to put the algorithm to practical due to the enormous amount of calculation required.

Moreover, Sungmin Kim etal.[26] introduced a Proof-of-Probability (*PoP*) method to solve the drawbacks in *PoW* and *PoS* methods, each node sorts the encrypted actual hash as well as a number of fake hash, and then the first node to decrypt actual hash creates block. Proof of Retrievability (*PoR*) [27] increases the ability to provide provable commitment protocols by considering bandwidth and retrievability. Proof of Ownership (*PoO*) [28] is considered with regard to various goods and rights in terms of receiving services. Proof of Importance (*PoI*) [29] uses a method that clusters nodes through transaction graph analysis, utilizing the transaction quantities and the balances of individual nodes as indicators, determining the importance of each node and designating the priority using hash computations to more significant nodes. Aggelos Kiayias etal.[30] presented "Ouroboros", the first blockchain protocol based on proof of stake with rigorous security guarantees. It offers qualitative efficiency advantages over blockchains based on proof of physical resources

(e.g., proof of work). Maria Borge etal.[31] proposed proof-of-personhood (*PoP*), a mechanism that binds physical entities to virtual identities in a way that enables accountability while preserving anonymity.

## 2.6 Application of AI in blockchain

To solve the problem that too much computing and energy resources consumed on mobile devices in the mining process. Nguyen et al.[32] developed an optimal auction based on deep learning for the edge resource allocation. They construct a multi-layer neural network architecture based on an analytical solution of the optimal auction. The neural networks first perform monotone transformations of the miners' bids. Then, they calculate allocation and conditional payment rules for the miners.

## 3. Proof of AI

*PoW* (Proof of Work) protocol consumes a large number of electricity, *PoS* (Proof of Stake) and *DPOS* (Delegated Proof of Stake) are both essentially a centralized voting agreement. To overcome the shortcomings of the above protocols, we present a new energy-saving consensus protocol *PoAI* (Proof of Artificial Intelligence) to ensure the decentralization and safety of a block chain system.

In this section, the crucial concepts of *PoAI* are introduced in Section 3.1, and the overall process of the protocol is proposed in Section 3.2. Considering the goal to run the blockchain system stably and safely under *PoAI*, we design a training method combined with the technology of the convolution neural network in Section 3.3.

### 3.1 Definitions and Concepts about average transaction number

- **Assumption**

The blockchain is a transaction records data chain and all data in the peer-to-peer network is shared by nodes. The distributed blockchain network need to satisfy decentralization, security, and fairness as much as possible. Because of the high network delay in point-to-point network, the sequence of transactions observed by different node cannot be completely consistent.

- **Definition 3.1 computing power ratio**

Bitcoin's most basic hash algorithm requires only the most violent swallowing and processing power. In terms of bitcoin's execution efficiency, the overall computing power $Pow$ of hardwares strictly follows the order of CPU<GPU<DSP<ASIC. In the face of complex tasks, the most flexible and efficient CPU has the lowest efficiency and performance power ratio because it sacrifices the throughput capacity. Computing power ratio is defined as the ratio of calculating force of node I to calculating force of the whole network.

- **Definition 3.2 average transaction number** $ATN$

$ATN$ serves as the main basis for node selection, which measures the average transaction number that a node gets during per timing circle comprehensively. It can be modeled as an evaluation function that takes node properties $CRT$、nature of the network $HCL$、safety elements $DAA$ as its independent variables.

- **Definition 3.3 super nodes**

Super nodes is these node with more powerful computational capability, less network latency, more mining equipments. These super nodes can be picked under a mechanism by *PoAI*.

- **Definition 3.4 random nodes**

Random nodes are nodes apart from super nodes, which guarantee the fairness of the network. Although average transaction number of these nodes are lower than super nodes, random nodes are qualified to join the node pool.

- **Definition 3.5 capability threshold**

The parameter $Thres$ is a boundary changed by the performance of the blockchain network dynamically. It is a standard of classification to divide the super nodes and the random nodes.

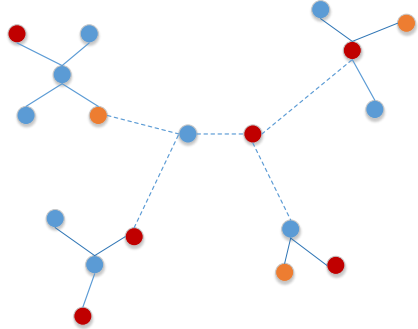

Fig.1 all nodes in a distributed peer-to-peer network, red nodes are super nodes, blue nodes are random nodes, and unknown nodes are the orange ones.

All nodes are divided into three classifications: super nodes, random nodes and unknown nodes.

$$N = Super \cup Random \cup Unknown$$

## 3.2 .Overview of Proof-of-Artificial Intelligence

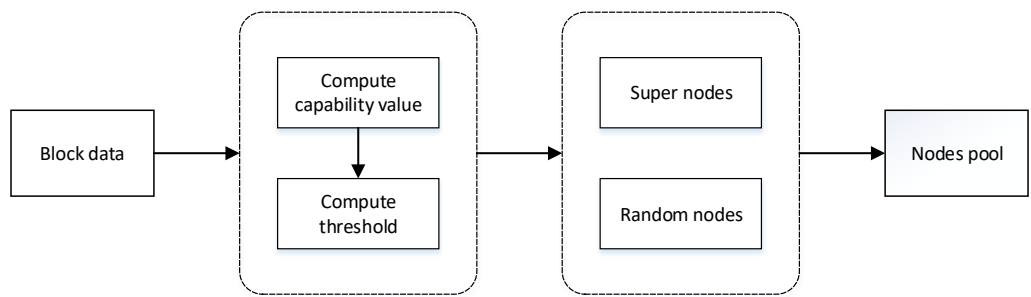

Fig.2 Proof-of-Artificial Intelligence block diagram.

A method for intelligently selecting an accounting node, relating to fields of blockchain, virtual currency and artificial intelligence, is provided, including steps of: (1) calculate the $AVN$ value of each node by Convolutional Neural Network (2) select the random nodes and the super nodes under a certain threshold value. Characteristic data of nodes are taken as the input information of **PoAI**, and then a node pool and the mining node are determined as outcomes.

On a premise of guaranteeing fairness, decentralization and security of the blockchain, problems of energy waste and low accounting efficiency due to mining conflict are solved. Specifically, a random distribution of accounting rights or mining rights is also adopted for avoiding hacker attacks. The process of constructing a node pool does not require human participation and does not require hash operation to compete computing power, which saves electricity, and ensures fairness and decentralization.

### 3.2.1 Compute Average Transaction Number of a Node

1. **Character Representation**

We extracts i-th node state by filling out the aptitude test questionnaires $M_i$. In this questionnaires, The average transaction number of each node is assessed on three indicators, including node properties $CRT$、nature of the network $HCL$、safety elements $DAA$. Primary and 9 secondary influence factors are shown as follows:

$$M = \{CRT, HCL, DAA\}$$
$$CRT = \{computing\ power\ ratio, online\ time, payoff\ \}$$
$$HCL = \{hop, connection\ number, latency\}$$
$$DAA = \{discarded\ probabiity, atracked\ probabiity, atract\ probability\}$$

2. **Node capability computation based on AlexNet**

Alex proposed the *AlexNet* network structure model in image classification and won the 2012 champion of ImageNet challenge. We propose a modified *AlexNet* to complete a capability assessment system. The essence is to fit the average transaction number ATN according to the input characteristic matrix $M$.

$$ATN = f(M) = f(CRT, HCL, DAA)$$

## 3.2.2 Pick Super nodes and Random nodes

All nodes are divided into three categories by Algorithm 2.

$$N_i = \{Super \cup Random \cup Unknown\}\ i = 1,2,\ldots, Num$$

**Algorithm 1**

1. Calculate ATN_sorted
2. Determine the maximum capacity of a node pool $Whole\_max$
3. Generate a random integer i (0.5 $Whole\_max$ <i<$Whole\_max$) as node number in a node pool $Node\_pool\_num$
4. Generate $Thres$ (<$Nodes\_pool\_num$) to determine the number of super nodes $Sup\_num$
5. Select the first $Sup\_num$ nodes as the Super nodes
6. Pick $RAD\_num$ random nodes

The *ATN* values can be obtained from the well-trained convolution neural network, and a ranking list *ATN_sorted* in descending order is easy to be calculated by stable *Merging sort algorithm* with low time and space complexity. One node is awarded as a super node as long as its rank is higher than $Sup\_num$. We pick R$AD\_num$ random nodes from the *ATN_sorted* except for Super nodes stochastically. Both super nodes and random nodes form the node pool. The mining node can be picked from the node pool based on a rotation mechanism.

A node pool consisting of super nodes and random nodes realizes fairness of the whole network since both powerful nodes and common nodes participate in transaction record. The mechanism of generating random nodes ensures that the distributed network is capable to resist external some attacks. As the number of nodes in the network increases, the node capacity of the node pool remains unchanged. It is a fact that only nodes in the node pool has the right to mine by turn. This consensus encourages nodes to enhance their personal strength $CRT$ and lower security risk $DAA$ as much as possible in blockchain.

## 3.3 Training $ATN$ by Convolution Neural Network

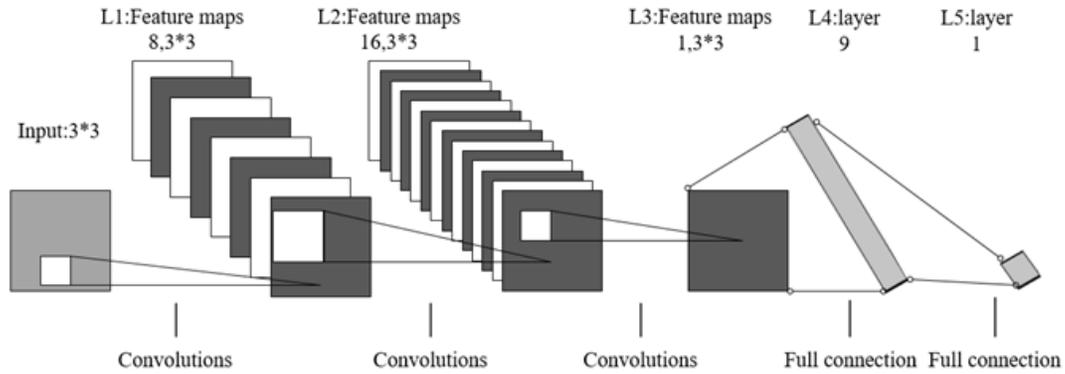

Fig.3 the block diagram modified *AlexNet*.

*AlexNet* with 5convolution layers, 3 connection layers, 60,000 parameters and 650,000 neurons realizes 1000 kinds of image classification. Activation function *ReLu*, pooling and normalization are operated after each convolution model. Our work makes the modified *AlexNet* predict the average transaction number of each nodes. Our model has 5 layers including three convolution layers, two full connection layers. In order to avoid over fitting problem and improve forecasting effect, we regularize the weights of each convolution layer by L2 norm.

State information of each node $M$ in a distributed network is taken as the dataset. The 2 dimensional matrix $M_i$ is the input of the convolution neural network, and the average number of becoming the mining node in a term is the efferent capacity label. After training our network, we can get average transaction number of the *i-th* node as long as $M_i$ is inputted.

## 4. Experiments

Based on the algorithm proposed in Section 3, the AICHAIN system is constructed which aims at providing a low-level blockchain controlled by AI and a DAE(digital asset entitlement) trading platform in AI ecology, as shown in Fig.3. AICHAIN will use AI to select the super nodes, and provide a more sophisticated AI application with the common block chain platform, and allows the data resource, application developer, operation platform and users to exchange their own resources on the de-centralized DAE platform.

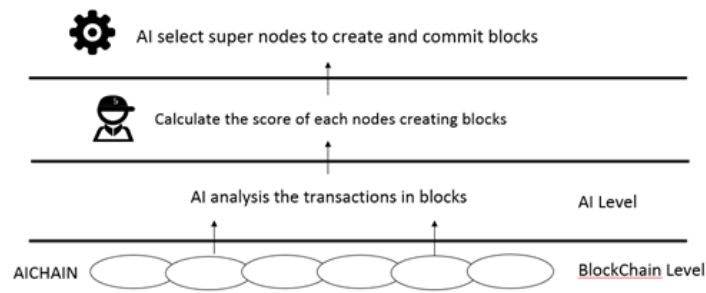

Fig.4 The low-level structure picture.

In order to capture the dataset used for our training model, we collect node state information from the

AICHAIN network including node properties $CRT$、nature of the network $HCL$、safety elements $DAA$. After labelled by hand, 0.1million samples are obtained which consists of the state information $M$ and the corresponding transaction number of each nodes. Our blockchain dataset contains node state information and the label of average transaction number. Part of the data are shown below.

| *Sequence number* | *computing power ratio* | *online time* | *payoff* | *hop* | *latency* |
|---|---|---|---|---|---|
| 1 | 0.12 | 1000 (s) | 5k s·BTC | 50 | 0.01 (s) |
| 2 | 0.22 | 650 (s) | 10k s·BTC | 125 | 0.001 (s) |
| 3 | 0.05 | 1200 (s) | 2.5k s·BTC | 90 | 0.001(s) |

**Table.1 Node state information $M$ and the label of average transaction number**

| *Sequence number* | *connection number* | *discarded probability* | *atracked probabiity* | *atract probability* | *average transaction number* |
|---|---|---|---|---|---|
| 1 | 0.2k | 0.04 | 0.002 | 0 | 65.2 |
| 2 | 1k | 0.03 | 0.001 | 0 | 125 |
| 3 | 0.9M | 0.12 | 0.001 | 0.001 | 7.5 |

Our method is implemented using the ***Keras*** framework based on deep learning. Training was performed on an Intel(R) Core(TM) i7-6700 CPU machine with 8GB memory.

We selected five of these factors to test our model. Fig.5 shows the variation of the average number of transactions as these factors changed, where the blue dots represents the real label of the average transaction number, and the red dots represents the values predicted using our model. In Fig.5 (a), the relationship between the average transaction and the computing power ratio are described. We can see that the average number of transactions shows an increasing trend with the increase of computing power ratio, which is consistent with the actual situation. Fig.5 (b) shows that as the payoff increases, more transactions the node will have. We can also observe in Fig.5(c), (d), (e) that the transaction number of each node is negatively correlated with the number of hops, the transmission latency and the attacked probability of each node. We can see that the variation trend of the predicted result is consistent with the variation of the real value. Fig.5 (f) represents the difference values between the predictions and the real labels about the five parameters (computing power ratio, payoff, hop, latency, attacked probability). In all case, the model we proposed consistently achieved the best results, and it shows that our model is reasonable.

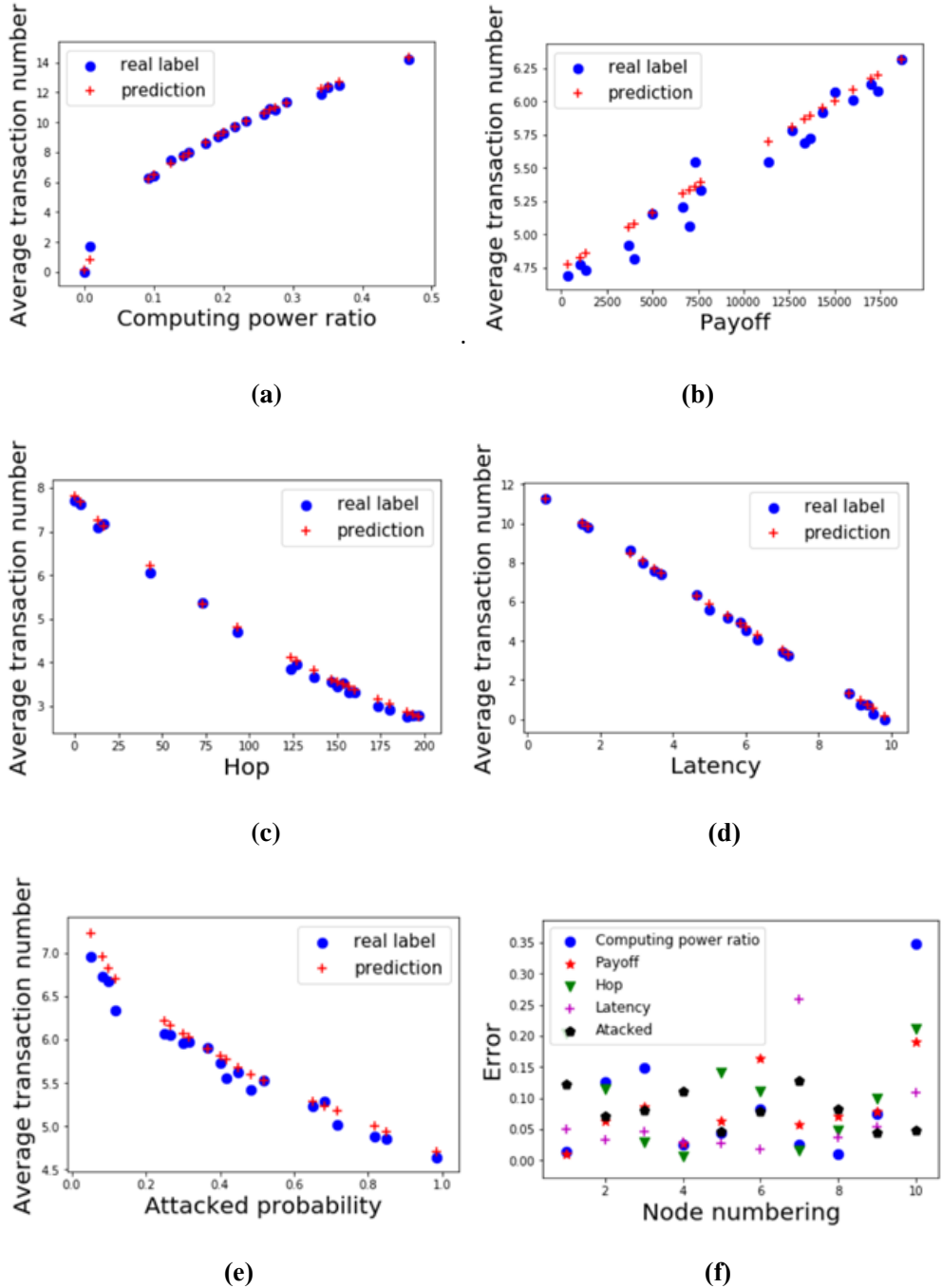

Fig.5 Average transaction number of different factors

In Fig.6, We compared the mining nodes of our method with the existence consensus protocols. Fig.6 (b) shows the mining nodes selected by our method, where the blue dots represents the super nodes and the red dots represents the random nodes. We can see that, through the method our proposed, some random nodes have been added to the mine, which increases the security of the network and providing opportunities for new participants. It can increases the fairness of the mechanism we proposed.

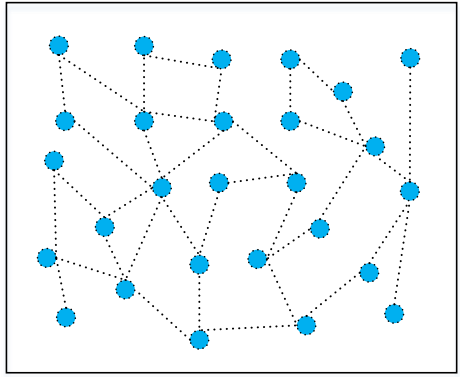

(a) Mining nodes selected by other protocols

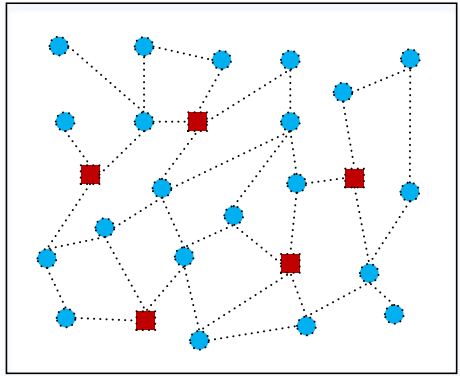

(b)  Mining nodes with random nodes selected by POAI

Fig.6 Comparison the mining nodes of our method with the existence consensus protocols.

The algorithm firstly merges the advantages of these three known mechanisms, namely, computing power ratio and payoff are regarded as the important factors affecting this node selection algorithm. In addition, we consider other factors that determine node performance, network structure, and node state, such as latency, hop, and attack probability. These factors mentioned above make the proposed algorithm more comprehensive and node selection is more fair, specifically embodied in decentralized. Different from the conventional consensus protocols, our algorithm does not rely on people to select nodes, which greatly saves resources and shortens the validation time. Meanwhile, random nodes improves the security of the proposed mechanism.

## 5. Conclusion

In this paper, we proposed a novel nodes selection algorithm based on AI technology that exploits nearly complementary information of each nodes, and relies on a particular designed convolutional neural network to reach the consensus. To ensure the decentralization and safety of the network, a dynamic threshold was employed to obtain the super nodes and the random nodes. Our algorithm avoided the complicated hash operation and redundant verification operation, which can be beneficial to save energy.

In our experiments, the proposed algorithm performs reasonably in aspects of computing power ratio, payoff, hop, latency and attacked probability, which combines the advantages of *PoW*, *PoS*, and *DPoS*. Furthermore, it compares favorably to these three state-of-the-art consensus frameworks, in terms of decentralization, network cost, security and the speed of transaction confirmation. Experimental results with our algorithm demonstrate that it can be adopted to every cryptocurrency and can be developed to a complete consensus protocol in the near future.

## 6. Acknowledgements

We express our thanks to the AICHAIN FOUNDATION LTD, Singapore and the MediaLab in UESTC, which supported this research work.